\documentclass[aps,prl,twocolumn,groupedaddress]{revtex4-1}

\usepackage{graphicx}

\begin{document}

\title{Spin crossover in ferropericlase from first-principles molecular dynamics}

\author{E. Holmstr\"om}
\email[]{e.holmstrom@ucl.ac.uk}

\author{L. Stixrude}

\affiliation{Department of Earth Sciences, University College London,
  Gower Street, London WC1E 6BT, UK}

\date{\today}

\begin{abstract}
Ferropericlase, (Mg,Fe)O, is the second-most abundant mineral of the
Earth's lower mantle. With increasing pressure, the Fe ions in the
material begin to collapse from a magnetic to non-magnetic spin
state. We present a finite-temperature first-principles phase diagram
of this spin crossover, finding a broad pressure range with coexisting
magnetic and non-magnetic ions due to favorable enthalpy of mixing of
the two. Furthermore, we find the electrical conductivity of the
mineral to reach semi-metallic values inside the Earth.
\end{abstract}


\maketitle

Ferropericlase, (Mg$_{1-x}$Fe$_x$)O, is an Fe-bearing transition metal
oxide that makes up some 20\% of the total volume of the Earth's lower
mantle~\cite{Lin2013}. Each Fe ion in this mineral assumes an
octahedral coordination environment, which leads to crystal field
splitting, {\it i.e.}, separation of the Fe $3d$ shell of electrons
into a higher-energy $e_g$ and a lower-energy $t_{2g}$ group. At low
pressure, the ground-state electronic configuration of Fe$^{2+}$ is a
high-spin state with four unpaired electrons giving a total spin of $S
= 2$. On compression, three effects come into play which ultimately
cause a spin transition or magnetic collapse to the low-spin state, $S
= 0$. Firstly, the crystal field splitting grows due to increased
overlap of the Fe and O valence orbitals, while secondly, the
electronic bands are broadened in energy due to increased confinement,
making the high-spin state increasingly
unfavorable~\cite{Cohen1997,Shulenburger2010}. Thirdly, the low-spin
state is favored by the smaller size of the low-spin Fe ion via the
$P\Delta V$ contribution to the free energy
~\cite{Persson2006,Crowhurst2008}.

Ever since the discovery of the spin transition in Fe$_{0.94}$O beyond
60~GPa at room temperature~\cite{Pasternak1997} and later in
ferropericlase between pressures of 50 to 70 GPa~\cite{Badro2003}, it
has become apparent that the phenomenon affects
mechanical~\cite{Lin2005,Fei2007,Crowhurst2008,Wentzcovitch2009,Wu2013},
compositional~\cite{Badro2003}, and electronic
properties~\cite{Goncharov2006,Lin2007}, and thus holds potentially
significant implications for the physics and chemistry of the
Earth. Interest in spin transitions is not, however, limited to
geoscience, with applications in, \textit{e.g.}, nanoclusters and thin
films showing great technological
potential~\cite{Bousseksou2011}. Experimental work utilizing x-ray
emission spectroscopy (XES)~\cite{Badro2003,Lin2005}, optical
spectroscopy~\cite{Goncharov2006}, M\"ossbauer spectroscopy
(MSB)~\cite{Kantor2006}, and equation-of-state (EOS) data gathered
from high-pressure x-ray diffraction experiments~\cite{Mao2011} has,
to date, probed the spin transition in ferropericlase up to pressures
of $P = 140$~GPa and temperatures of $T=2000$~K. On the theoretical
side, approaches based on analytical mean-field
theory~\cite{Sturhahn2005} and static first-principles
calculations~\cite{Tsuchiya2006} augmented by quasi-harmonic phonon
computations~\cite{Wentzcovitch2009} have treated the spin transition
as in fact a smooth \textit{spin crossover}, an approach consistent
with published experiments. This crossover proceeds, with increasing
pressure, from all Fe ions assuming the high-spin state, through to a
mixed-spin phase with coexisting high-spin and low-spin ions, to
eventually all ions assuming the low-spin state.

Previous theoretical work has been based on static calculations and
has assumed that the mixed-spin state is stabilized entropically,
yielding a narrow crossover at low temperatures that disagrees with
experiment. Moreover, experiment and theory have not explored
geophysically important properties such as the band structure and
electrical conductivity. In this Letter, we take a different approach
that combines first-principles molecular dynamics with free-energy
minimization to simulate the high-temperature properties of the spin
crossover directly. Our results reveal a new physical picture of the
crossover, where the mixed-spin phase is stabilized through enthalpy
rather than entropy, giving a finite broadness for the crossover even
at vanishing temperatures. Additionally, we predict the EOS up to the
conditions at the base of Earth's mantle (140~GPa, 4000~K), and find
that the electrical conductivity of ferropericlase reaches
semi-metallic values at the bottom of the lower mantle, with
significant geophysical implications.

Our simulation setup is built on molecular dynamics (MD) simulations
within density functional theory (DFT), as implemented in the VASP
package~\cite{VASP}. We consider a cubic simulation cell of 64 atoms
with periodic boundary conditions, adopting an Fe concentration of
$x_{Fe} = 25$\%, with the Fe ions arranged in a regular superlattice
with two nearest-neighbor distances between any two neighboring Fe
ions. In order to obtain an efficient simulation setup with
well-converged values for internal energy and pressure (within
5~meV/atom and 0.2 GPa), we sample the Brillouin zone at the
Baldereschi point~\cite{Baldereschi1973,Martin2008} for a lattice of
simple cubic symmetry and use a planewave cutoff energy of 500~eV. The
projector-augmented wave method is used to avoid explicit calculation
of the core electron orbitals. To decide on the best feasible
approximation to the exchange and correlation part of the total energy
functional, we compared the EOS from conjugate-gradient relaxed static
calculations to experiment at 300~K, using the local-density
approximation (LDA) and two different generalized-gradient
approximations (GGA), PBE~\cite{PBE} and PBEsol~\cite{PBEsol}. Of
these functionals, PBEsol proved clearly superior.

Unfortunately, the PBEsol functional fails to fully capture the strong
correlation between the $3d$ electrons of the Fe ions, which is
manifested as a spin transition pressure of only 18~GPa for $x_{Fe} =
3.125$\%, whereas the experimental estimate is closer to
50~GPa~\cite{Goncharov2006}. As meta-GGA type
functionals~\cite{M06L,TPSS,RTPSS,Sun2011} that we tested brought no
alleviation to this problem, and as hybrid functional calculations
utilizing the exact Fock exchange of the DFT system of quasi-electrons
are computationally too demanding for MD, we use the $+U$
method~\cite{Dudarev1998} to approximate the forementioned correlation
effects. Based on our calculations on the dependence of the spin
transition pressure on $U$, an empirical estimate for $U$ from optical
spectroscopy data~\cite{Goncharov2006}, as well as our own hybrid
functional~\cite{HSE06} computations, we settled on
$U-J = 2.5$~eV. We then performed our PBEsol+$U$ MD simulations in the
$NVT$ ensemble using the Nos\'e-Hoover thermostat. Each simulation was
run with a timestep of 1.0~fs for a total of 10~ps to reach thermal
equilibrium, followed by 10~ps over which all physical time averages
were computed. A total of three isotherms, $T =$~2000, 3000, and 4000
K were simulated for compressions that result in pressures of
approximately 0 to 200~GPa, to encompass existing experimental data
and the conditions of the lower mantle of the Earth. These dynamic
computations were complemented with a set of static calculations,
where the crystal structure was relaxed using conjugate gradients.

In order to capture the continuous character of the spin crossover and
to thus produce a first-principles phase diagram of the phenomenon, we
minimize the Gibbs free energy $\Delta G(P,T,f) = G(P,T,f) - G(P,T,0)$
at each $P$ and $T$ with respect to $f$, the fraction of Fe ions in
the high-spin state. As we find a vanishingly small amount of
intermediate spin ($S = 1$) Fe in our simulations, we define $f \equiv
\langle \mu_{Fe} \rangle / \langle \mu^{HS}_{Fe} \rangle$, where
$\mu_{Fe}$ and $\mu_{Fe}^{HS}$ denote the Fe magnetic moment and the
same when all Fe ions are in the high-spin state, respectively, and
$\langle \rangle$ denotes an average over Fe ions and time. To map
$\Delta G(P,T,f) = \Delta H - T \Delta S$ as a function of $f$, we
perform constrained-moment and free-moment calculations, the former
producing a low-spin ($f = 0.0$) and high-spin ($f = 1.0$) phase and
the latter producing two mixed-spin phases along each isotherm. The
enthalpy $H$ of a given phase $f$ is obtained directly from the MD
simulation, as is the electronic contribution to the entropy
$S_{el}$~\cite{Mermin1965,Kresse1998} (we set the electronic
temperature equal to the ionic temperature). The vibrational entropy
$S_{vib}$ and entropy $S_{conf}$ due to site-switching of high spins
and low spins we evaluate through the method of thermodynamic
integration~\cite{Vocadlo2002,Gillan2006}. The last contribution to
the entropy, $S_{mag}$ due to the fully disordered paramagnetic state
of the moments above the N\'eel temperature
of~$\sim$500~K~\cite{Speziale2005}, we compute from the
expression~\cite{Grimvall1989} $S_{mag} = k_B\sum_i\ln(\mu_i+1)$,
where $\mu_i$ is the total magnetic moment of Fe ion $i$, and $k_B$ is
the Boltzmann constant. We thus obtain $\Delta G(P,T,f)$ at four
values of $f$ for each $P$ and $T$, and to find the equilibrium $f$,
we interpolate and minimize $\Delta G(P,T,f)$ with respect to $f$
using a free second-order polynomial~\cite{SMHolmstrom2014}.

%
%
\begin{figure}
\includegraphics[width=0.5\textwidth]{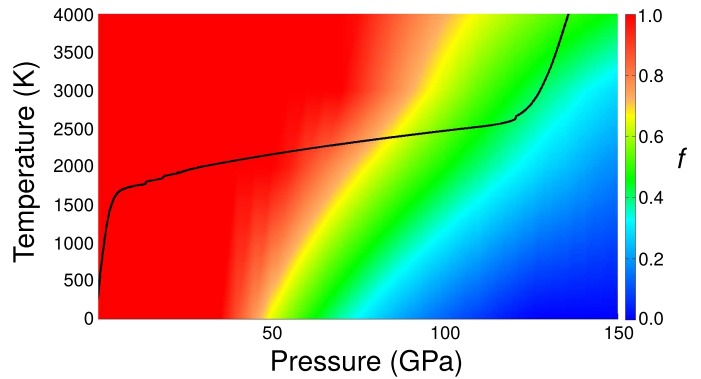}
\caption{Our first-principles phase diagram of the spin crossover in
  ferropericlase. The black line is a geotherm
  from~Ref.~\cite{Stixrude2009}.\label{fig:phase_diagram}}
\end{figure}

\begin{figure*}
\includegraphics[width=1.00\textwidth]{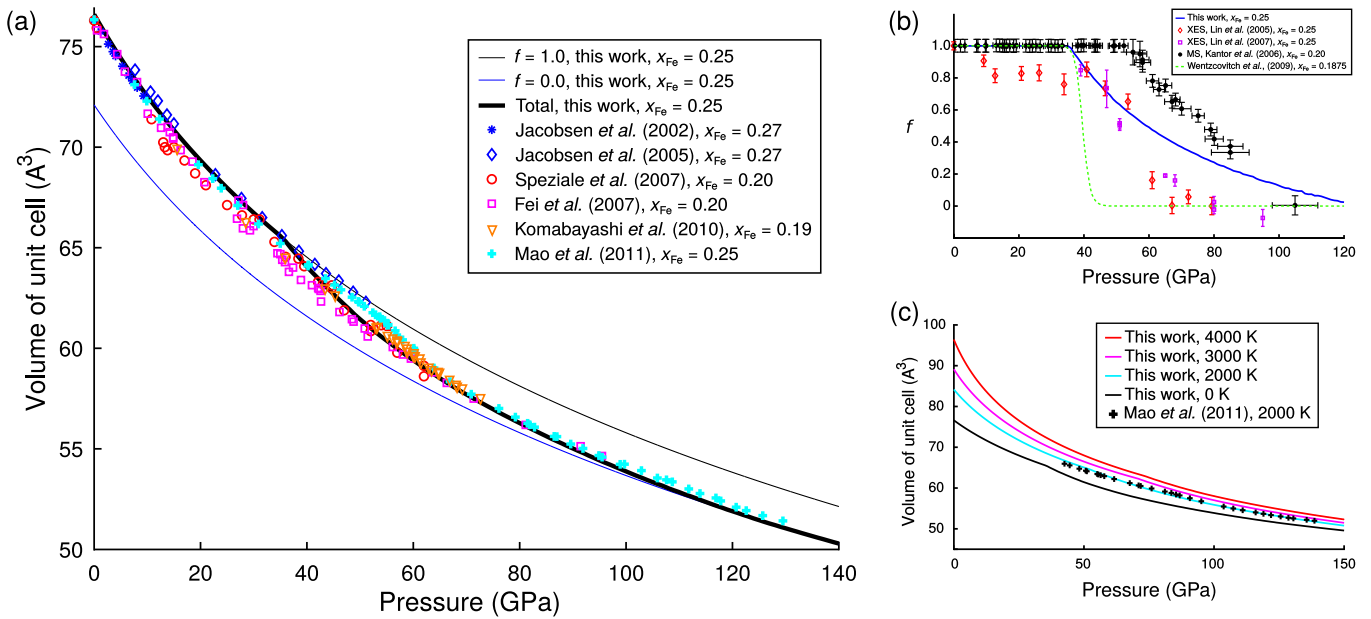}
\caption{a) The EOS of ferropericlase at static conditions compared to
  experimental data gathered at
  300~K~\cite{Jacobsen2002,Jacobsen2005,Speziale2007,Fei2007,Komabayashi2010,Mao2011}. We
  construct the total EOS by interpolating $V=V(P,f)$ linearly between
  the four spin phases $f$, where the EOS for each phase is a fit to
  the third order Birch-Murnaghan EOS~\cite{Birch1986}. b) Our static
  result for $f$ compared to XES and MSB
  data~\cite{Lin2005,Lin2007,Kantor2006} as well as previous
  computational results~\cite{Wentzcovitch2009} at 300~K. c) Our
  thermal EOS for all simulated temperatures along with experiment at
  2000~K~\cite{Mao2011}.}
\label{fig:eos_and_xes}
\end{figure*}

The resulting phase diagram for the spin crossover is presented in
Fig.~\ref{fig:phase_diagram}. Strikingly, we find a broad pressure
interval of coexisting high-spin and low-spin ions at all
temperatures, even at fully static conditions ($T = 0$~K in the phase
diagram). Another interesting feature of the phase diagram is the weak
temperature-dependence of the stability field of the mixed-spin phase
up to $\sim$3000~K. The shape of our phase diagram is thus
fundamentally different from previous theoretical
work~\cite{Sturhahn2005,Tsuchiya2006,Wentzcovitch2009}, where the
mixed-spin phase was stabilized through an ideal mixing entropy,
resulting in a completely sharp spin transition at $T = 0$~K. We
predict $f \approx 0.5$ at the core-mantle boundary, also at odds with
previous computations, which have found significantly smaller
high-spin fractions. Comparison of our static results to existing
experimental EOS, XES, and MSB data at 300~K shows overall good
agreement (Fig.~\ref{fig:eos_and_xes}a,b). Previous computations show
a much narrower crossover than EOS, XES, and MSB data, and our present
results. Our results for the EOS at all simulated temperatures are
presented in Fig.~\ref{fig:eos_and_xes}c, displaying good agreement
with experimental high-temperature data.

%
%

\begin{figure}
\includegraphics[width=0.5\textwidth]{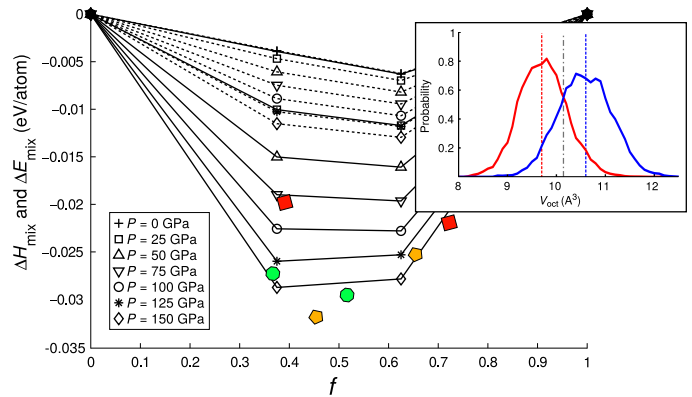}
\caption{Enthalpy (solid lines) and internal energy (dashed lines) of
  mixing of high-spin and low-spin Fe at static conditions, along with
  enthalpy of mixing at 2000~K and 91 and 64 GPa (green octagons, left
  to right, respectively), 3000~K and 98 and 73 GPa (orange
  pentagons), and 4000~K and 106 and 81 GPa (red squares). Inset:
  Distribution of octahedral volumes at 2000~K and $f = 0.50$ for Fe-O
  octahedra of high-spin Fe (blue solid line) and low-spin Fe (red
  solid line). The corresponding static results are shown by the
  dashed vertical lines. The black dash-dotted line denotes octahedral
  volume for $f = 1.0$ and 0.0 for static and 2000~K
  results.\label{fig:enthalpy_of_mixing_at_zero_K}}
\end{figure}

The finite width of the spin crossover even at vanishing temperatures
is due to favorable enthalpy of mixing $\Delta H_{mix}$ of the
high-spin and low-spin ions
(Fig.~\ref{fig:enthalpy_of_mixing_at_zero_K}). We trace the favorable
$\Delta H_{mix}$ to packing considerations arising from the volumes of
alternating high-spin and low-spin Fe-O octahedra. Due to occupation
of $e_g$ orbitals, the high-spin octahedron is larger than the
low-spin octahedron, and the Mg-O octahedron is intermediate in
size. When high-spin and low-spin Fe in (Mg,Fe)O are brought close
together, the system can exploit the willingness of high-spin Fe-O
octahedra to expand and their neighboring low-spin octahedra to
contract with respect to the MgO crystal (see inset of
Fig.~\ref{fig:enthalpy_of_mixing_at_zero_K}), resulting in lower
internal energy and forces and hence lower enthalpy than expected from
ideal mixing of high-spin and low-spin Fe. These results are
consistent with indications that Fe ions in (Mg,Fe)O tend to cluster
at high pressure~\cite{Kantor2009}. Our finding is, however, in sharp
contrast to previous computational work on the spin
crossover~\cite{Tsuchiya2006,Wentzcovitch2009}, where the high-spin
and low-spin ions have been assumed to form an ideal solid
solution. The favorable $\Delta H_{mix}$ that stabilizes the
mixed-spin phase in our static simulations persists in the dynamical
simulations (Fig.~\ref{fig:enthalpy_of_mixing_at_zero_K}).

%
%

Increasing temperature favors the high-spin state because of the
favorable contribution to the free energy from the $S_{mag}$ term and
to a lesser extent the $S_{vib}$ term. Over the range of temperatures
that we have considered, the magnetic entropy dominates over the
electronic entropy, which favors the low-spin state. The mixed-spin
phase region becomes slightly broader with increasing temperature due
to the increase in $S_{conf}$ with increasing temperature. $S_{conf}$
increases with temperature due to the increased occurrence of
spontaneous interchanges of high-spin and low-spin moments among the
Fe sites in the mixed-spin phase.

We find that the vibrational entropy is greater for high-spin ions
than for low-spin ions. This we relate to the shape of the valence
charge density of the Fe ion in the high-spin state which, considering
a sole Fe ion in MgO at static conditions and zero pressure, results
in a less symmetrical Fe-O octahedron (two axes expanding, one
contracting) than for the low-spin state (all axes contracting
uniformly). This underlying differential distortion, as quantified in
our $NVT$ simulations by the difference in octahedral quadratic
elongation~\cite{Robinson1971} between the high-spin and low-spin
phases, persists at finite temperature, leading to larger mean squared
displacements of the high-spin Fe ions and hence larger $TS_{vib}$ in
the corresponding phase~\cite{SMHolmstrom2014}.

The partitioning of Fe between ferropericlase and the major lower
mantle phase (Mg,Fe)SiO$_\textrm{3}$ perovskite has important
implications for understanding the structure, dynamics, and
geochemistry of the Earth's lower mantle~\cite{Badro2003}. We assess
the effect of the spin transition on the partitioning by computing the
ratio $\ln(K_{f}/K_{1.0})$, where $K_{f}$ is the partition coefficient
assuming the equilibrium $f$, and $K_{1.0}$ is the coefficient
assuming $f = 1.0$. Assuming no subsequent spin transition in the
perovskite, we find $\ln(K_{f}/K_{1.0})$ to lie approximately in the
range 0 to 1.5 along the geotherm~\cite{SMHolmstrom2014}, much less
than the value of $\sim$10 estimated by Badro \textit{et
  al.}~\cite{Badro2003}. Our much more moderate result for the effect
of the spin transition on the partitioning appears in better agreement
with the relatively weak pressure-dependence of $K_f$ found in
experiment~\cite{Lin2013}.

The electrical conductivity of the lower mantle is important for
understanding anomalies in Earth's rotation via electromagnetic
coupling of mantle and the underlying core, and the relationship
between observations of the geomagnetic field and its source through
the filter of a potentially conductive mantle. However, no
measurements or \textit{ab initio} predictions of the conductivity of
ferropericlase at conditions of the deep lower mantle are available.
Using the Kubo-Greenwood method to compute the electronic component of
$\sigma$ as implemented in VASP~\cite{Desjarlais2002,Pozzo2011}, we
find $\sigma=4.0 \pm 0.4 \times 10^4$~S/m at conditions close to the
bottom of the mantle ($P = 136 $~GPa, $T = 4000$~K), approximately
half the recently obtained value of $9 \times 10^4$~S/m for FeO in
similar conditions~\cite{Ohta2012}, consistent with the experimental
result that $\sigma$ increases with Fe
concentration~\cite{Li1990}. From the electronic density of states, it
is evident that the metallization of the mineral from its initially
insulating state is due to the $3d$ electrons of the Fe ions forming
broad bands that lead to a significant density of states at the Fermi
level, an effect due to both pressure and temperature. The spin
crossover itself serves to increase $\sigma$, as an increase in the
concentration of low-spin Fe implies increased density of states near
the Fermi level of the crystal~\cite{SMHolmstrom2014}.

The predicted semi-metallic value of electrical conductivity of
ferropericlase at the core-mantle boundary might be invoked to explain
the highly conductive layer in this region inferred from observations
of the planet's nutations and anomalies
therein~\cite{Buffett1992,Buffett2000}. Assuming the pyrolitic volume
fraction of 20\% for ferropericlase in the lower mantle, the presently
obtained electrical conductivity for the mineral, and taking the
surrounding perovskite phase to be insulating, the Hashin-Shtrikman
minimum-maximum bounds~\cite{Berryman1995} for the conductivity of the
mixture are zero and $5.7 \times 10^3$~S/m, respectively. Adopting the
maximum value and the half-way value, a simple calculation shows that
respectively 18 or 35~km of lower-mantle material is enough to give
the required minimum conductance of $10^8$ S to explain the nutation
observations. A more highly conductive mantle than previously assumed
may also require revision of the interpretation of surface
measurements of the Earth's magnetic field.

This research was funded by the European Research Council Advanced
Grant MoltenEarth. Calculations were performed on the Iridis computing
cluster partly owned by University College London, and HECToR and
ARCHER of the UK national high-performance computing service. The
authors acknowledge many valuable comments from R. Jeanloz, as well as
fruitful suggestions from R. E. Cohen and B. Militzer.

\end{document}


\section{Supplemental Material to ``Spin crossover in ferropericlase from first-principles molecular dynamics''}

\subsection{Details of the free-energy computations}

To compute the full Gibbs free energy for a given isotherm for the
high-spin phase ($f = 1.0$), low-spin phase ($f = 0.0$), and for $f$
intermediate between 0.0 and 1.0, we start by constructing the
pressure-volume equation of state (EOS) for each $f$. For $f = 1.0$
and $0.0$, we perform the MD simulations in the canonical $NVT$
ensemble over a range of compressions in such a way that the total
number of unpaired electrons in the supercell is constrained to
correspond to either four or zero per Fe ion, respectively. We fit a
third-order Birch-Murnaghan EOS~\cite{Birch1986} to the resulting
pressure-volume points.

\begin{figure}
\includegraphics[width=0.8\textwidth]{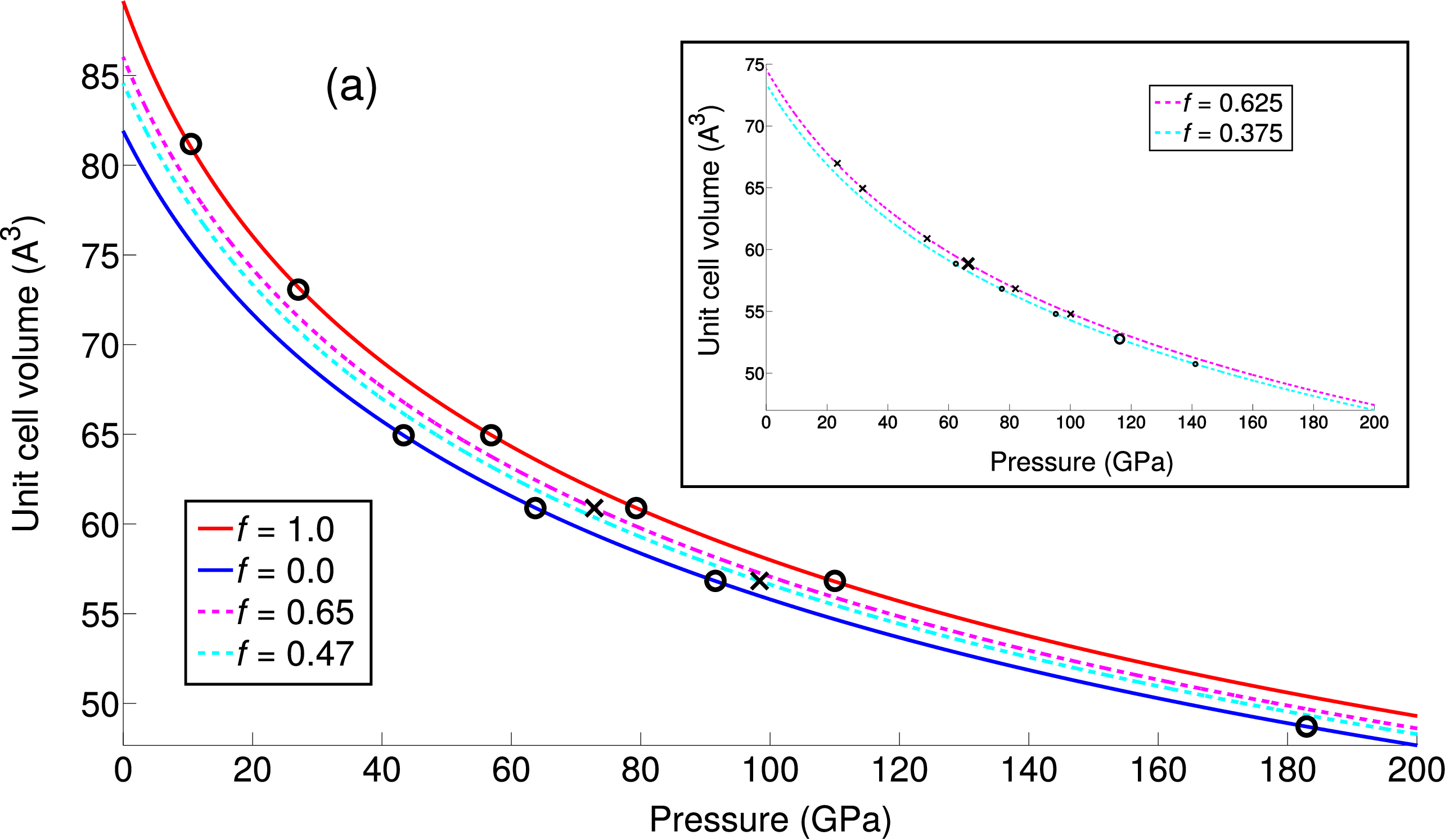}
\caption{The final result for the EOS at each value of $f$, this
  example being for $T = 3000$~K. The EOS for $f = 1.0$ and $f = 0.0$
  are fitted to data points from simulations with constrained moments
  (marked by circles). The EOS for $f = 0.65$ and $f = 0.47$ are
  constructed starting from free-moment simulation points (marked by
  crosses). Inset: Explicit test of the method used to construct the
  EOS for $f$ intermediate between 0.0 and 1.0, at static
  conditions. The large symbols indicate points from which the EOS
  were constructed, small symbols indicate points calculated
  explicitly after the construction of the EOS.\label{fig:eos_panel}}
\end{figure}

\begin{figure}
\includegraphics[width=1.00\textwidth]{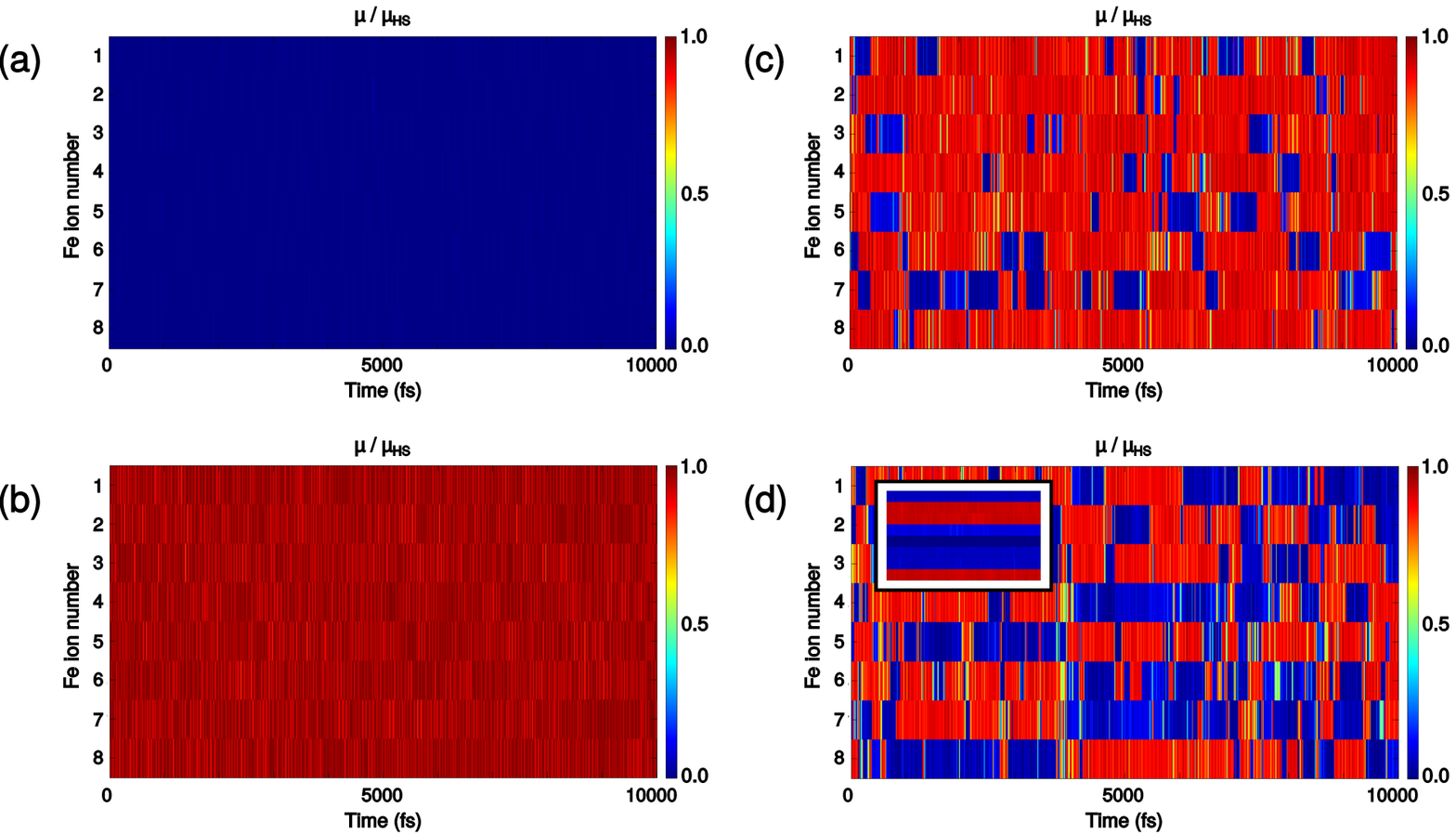}
\caption{Magnetic moment of each Fe ion in the 64-ion supercell as a
  function of time after the initial 10~ps equilibration, in units of
  the high-spin Fe moment. Data are shown for $f = 0.0$ (a), $f = 1.0$
  (b), $f = 0.76$ (c), and $f = 0.49$ (d) at $T = 4000$~K. The inset
  in (d) is the corresponding data for $f = 0.39$ at $T =
  2000$~K. \label{fig:magmom_vs_time_panel}}
\end{figure}

In order to construct the EOS for $f$ intermediate between 0.0 and
1.0, we first perform a set of free-moment MD simulations,
\textit{i.e.}, simulations where the number of unpaired electrons is
not constrained, using various compressions. Each such run produces a
certain average value for $f$. We choose two such pressure-volume
points, each with a different $f$ intermediate between 0.0 and 1.0, to
represent a point along the EOS of ferropericlase at each of these two
$f$, respectively (see Fig.~\ref{fig:eos_panel}a). We compute the
volume for each such intermediate $f$ at all other pressures via
\begin{equation} V(P,f) = V(P,0) + g(f)\Delta V(P),\end{equation}
where $\Delta V(P) = V(P,1) - V(P,0)$, and $g(f) = [V(P,f) - V(P,0)] /
\Delta V(P)$ is assumed to be constant and equal to the value computed
directly from our free-moment simulation for each $f$. The inset of
Fig.~\ref{fig:eos_panel} illustrates that this is a good approximation
and a sensible ansatz for computing the EOS of mixed-spin phases over
the entire pressure range of our calculations. Examples of the
magnetic structure for each $f$ are shown in
Fig.~\ref{fig:magmom_vs_time_panel}.

%
%

%
%

\begin{figure}
\includegraphics[width=0.85\textwidth]{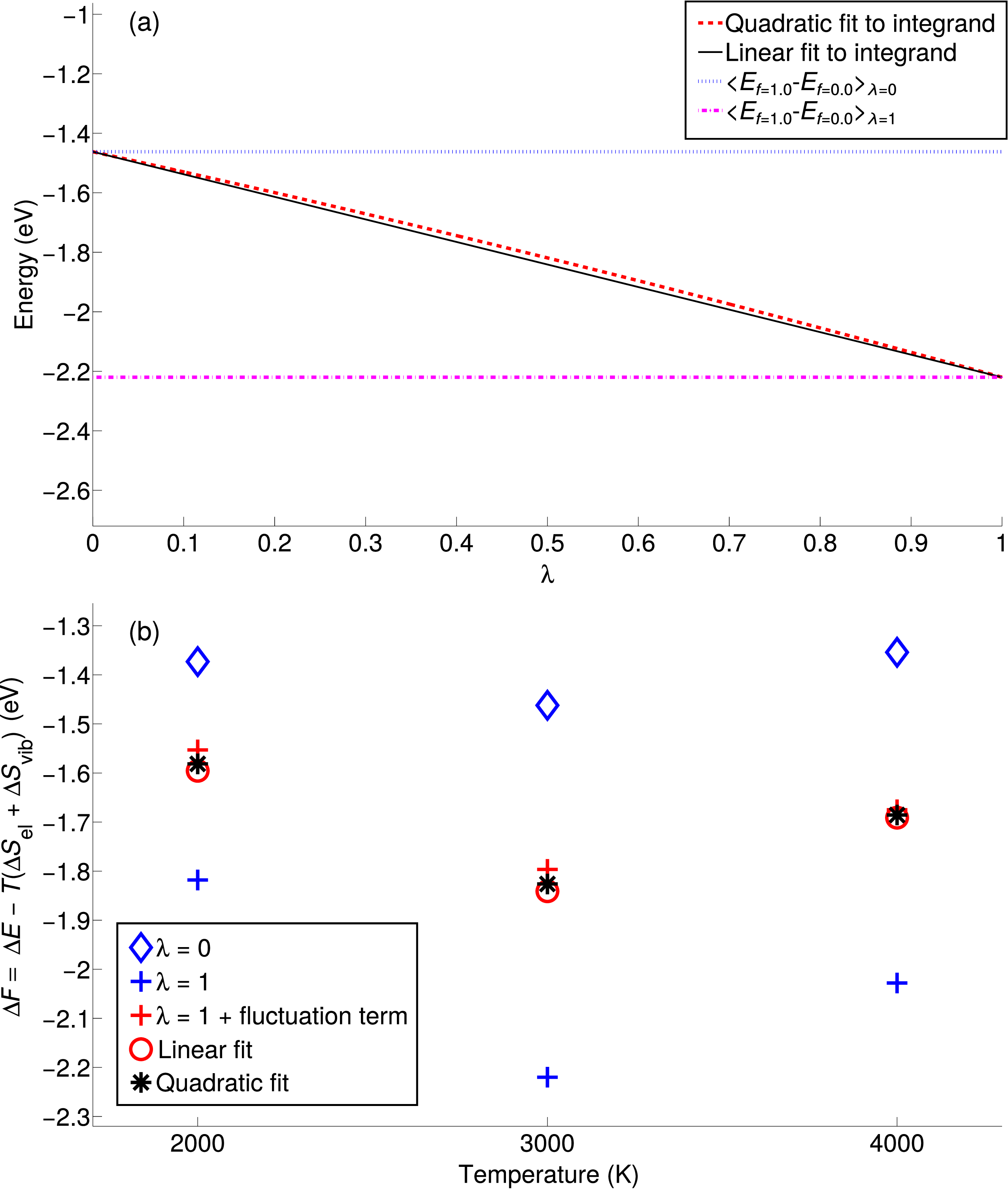}
\caption{a) The integrand of Eq.~\ref{eq:deltaF} at $T = 3000$~K at a
  lattice constant of 3.93~\AA~for $f=1.0$, using the approximation of
  constraining the integrand only at $\lambda = 0$ and 1 (linear fit)
  and the approximation of constraining additionally the slope of the
  integrand at $\lambda = 1$ (quadratic fit). The values of the
  integrand at $\lambda = 0$ and 1 are shown for reference. b) The
  value of the integral of Eq.~\ref{eq:deltaF} for $f=1.0$ for all
  isotherms and different levels of approximation. Using only $\lambda
  = 0$ or 1 results in overestimation or underestimation of $\Delta
  F$, respectively. The linear and quadratic fit agree well with the
  approximation of using $\lambda = 1$ and the fluctuation term of
  Eq.~\ref{eq:fluctuation}, implying good precision for the chosen
  approximations.\label{fig:ti_panel}}
\end{figure}

From the EOS for each value of $f$, we compute the Helmholtz free
energy $F = E - T(S_{el} + S_{vib} + S_{conf})$ by integrating the
thermodynamic identity $P = -(\partial F / \partial V)_T$. At this
stage, the free energy of each spin phase $f$ relative to the other
phases is unknown. To resolve this, we employ adiabatic switching
within the Kirkwood coupling scheme to compute the free energy of a
given phase $f$ relative to $f = 0.0$. The joint potential energy
function for switching from $f = 0.0$ to an arbitrary value of $f$
with the coupling parameter $\lambda$ going from 0 to 1 is
\begin{equation} E(\lambda) = E_{f=0.0} + \lambda (E_{f} - E_{f=0.0}). \end{equation}
The free energy difference between $f = 0.0$ and a given $f$ is
computed as the integral
\begin{equation} \Delta F = F_f - F_{f=0.0} = \int_0^1 d\lambda \langle E_{f} - E_{f=0.0} \rangle_{\lambda}, \label{eq:deltaF}\end{equation}
For $f = 1.0$, we use $\lambda = 0$ and 1 to constrain the integrand
at the end points. Additionally, we employ the fluctuation term
derived from thermodynamic perturbation
theory~\cite{Vocadlo2002,Gillan2006}
\begin{equation} \frac{\langle (E_{f=1.0} - E_{f=0.0})^2 \rangle_{\lambda=1} - \langle E_{f=1.0} - E_{f=0.0} \rangle^2_{\lambda=1}}{2k_BT}, \label{eq:fluctuation} \end{equation}
which, from simple geometrical arguments, is equal to $-1/2 \times
(d\langle E_{f=1.0} - E_{f=0.0} \rangle /d\lambda)_{\lambda = 1}$,
provided the perturbation term of Eq.~\ref{eq:fluctuation} is small
(we find 4 to 13 meV/atom throughout all calculations). We thus
constrain the slope of the integrand at $\lambda = 1$, and using the
three constraints, take a fully constrained second-order polynomial to
model the integrand. For $f$ intermediate between 0.0 and 1.0, we use
$\lambda = 1$ and the fluctuation term of Eq.~\ref{eq:fluctuation}, as
in~\cite{Vocadlo2002}. In Fig.~\ref{fig:ti_panel}, we present an
example of the integrand and an assessment of the relative precision
of the approximations detailed above, showing that overall the method
looks sufficiently precise for the present purposes.

%
%

\begin{figure}
\includegraphics[width=0.85\textwidth]{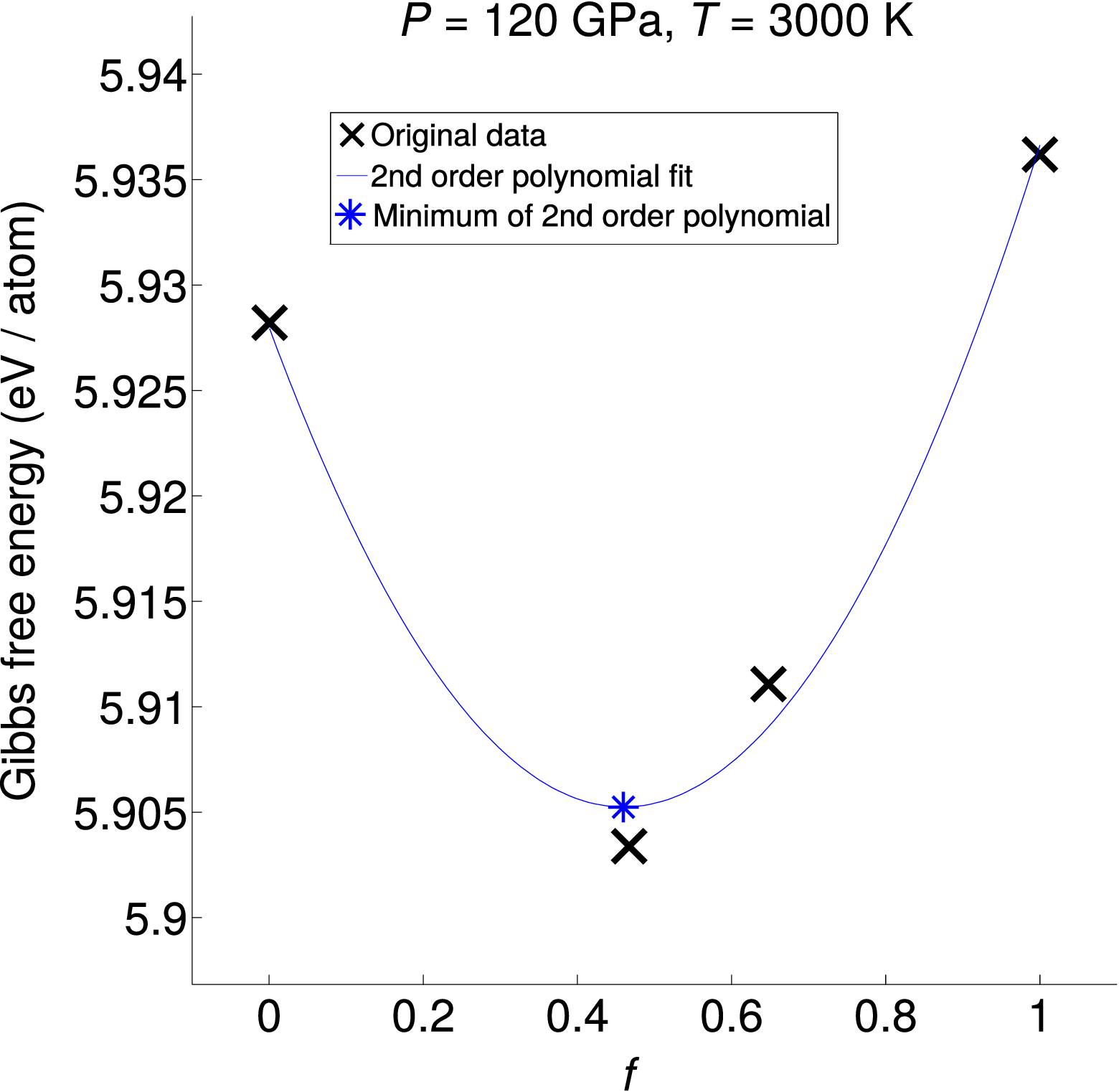}
\caption{Finding the equilibrium value of $f$ at $T = 3000$~K and $P =
  120$~GPa through second-order polynomial interpolation and
  minimization.\label{fig:interpolating_panel}}
\end{figure}

Finally, we add on the magnetic entropy $S_{mag}$ to the free energy
of each spin phase $f$, and move on to analyze the results in the
$NPT$ ensemble, \textit{i.e.}, in terms of Gibbs free energy. To find
the equilibrium high-spin fraction at each pressure and temperature,
we interpolate $\Delta G(P,T,f) = G(P,T,f) - G(P,T,0)$ with a free
second-order polynomial between the four values of $f$ and find the
minimum of that polynomial in the interval $f \in [0,1]$ (See
Fig.~\ref{fig:interpolating_panel} for an example). We consider this
method of interpolation a more neutral choice than using a third-order
polynomial or natural splines, both of which seemed to produce
spurious and artificial features into the interpolation.

%
%

\subsection{Explaining the vibrational entropy results}

\begin{figure}
\includegraphics[width=0.95\textwidth]{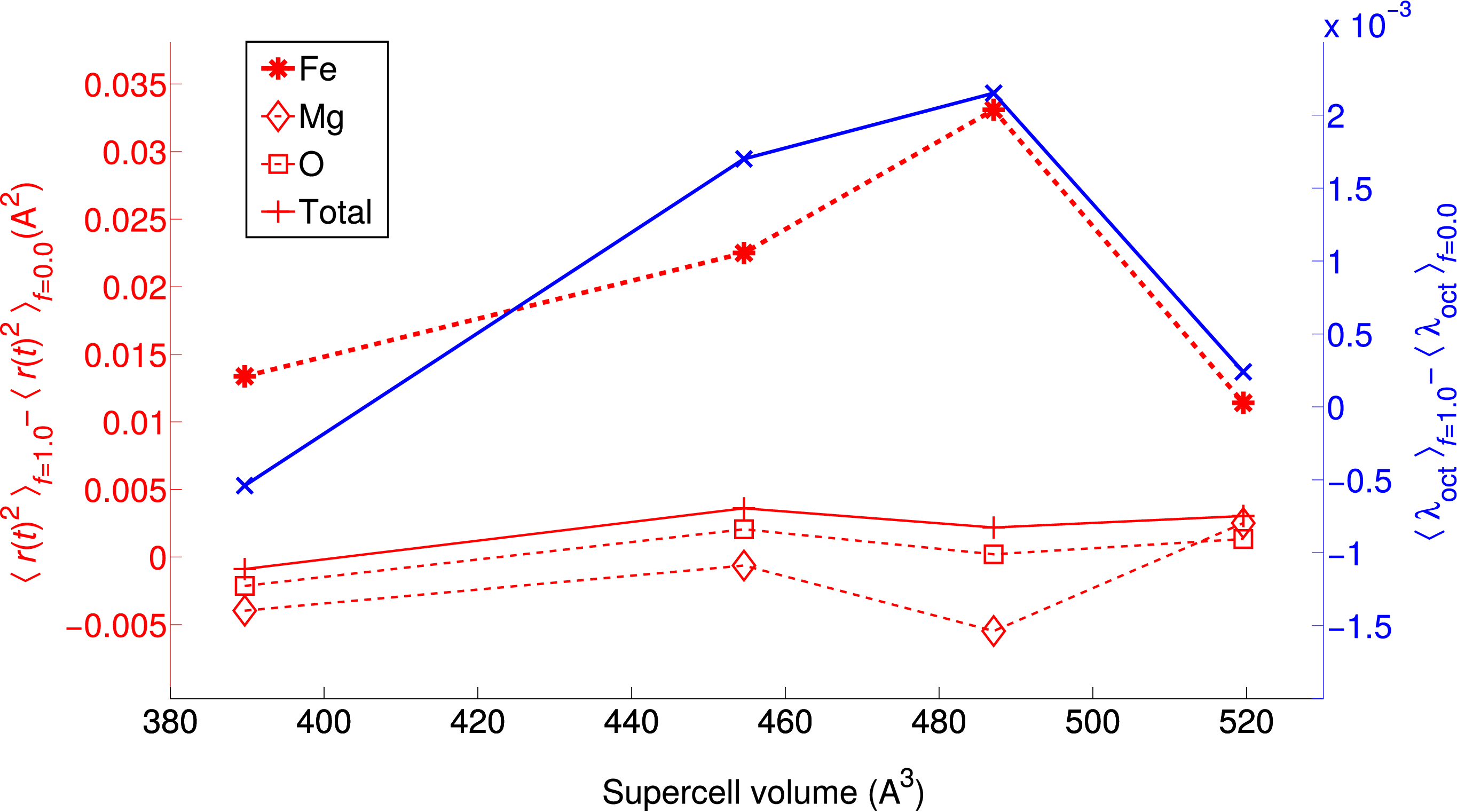}
\caption{Explaining the results for the vibrational entropy difference
  between $f=1.0$ and $f=0.0$, this example being at $T = 3000$~K. The
  difference between the two phases in mean squared displacements
  $\langle r(t)^2 \rangle$ of the Fe ions with respect to their mean
  positions is correlated with the difference in octahedral
  elongation, \textit{i.e.}, the degree of distortion of the Fe-O
  octahedra. The difference in vibrational entropy $TS_{vib}$ between
  $f = 1.0$ and $f = 0.0$ is of the order of 15~meV/atom here. The
  mean octahedral volume is the same for both phases in these $NVT$
  simulations.
\label{fig:svib_panel}}
\end{figure}

Populating the spin-up and spin-down $t_{2g}$ and $e_{g}$ states of a
single Fe ion within an MgO lattice unevenly to produce the high-spin
configuration entails a valence electron density around the ion which
results in a non-uniformly distorted Fe-O octahedron, contrary to the
situation in the low-spin state. We quantify the distortion of the
Fe-O octahedra in the dynamic simulations by using the mean octahedral
quadratic elongation $\langle \lambda_{oct}
\rangle$~\cite{Robinson1971}, and find Fe-O octahedra at $f = 1.0$ to
be generally more distorted than their counterparts at $f = 0.0$.  In
Fig.~\ref{fig:svib_panel} we show that the difference in mean squared
displacements of the Fe ions between $f = 1.0$ and 0.0 are correlated
with the difference in $\langle \lambda_{oct} \rangle$ between the two
phases. This implies that the ionic network at $f = 1.0$,
\textit{i.e.}, with all Fe ions in the high-spin state, is more
distorted than at $f = 0.0$, with all Fe ions in the low-spin state,
leading to larger nuclear excursions and hence the observed larger
vibrational entropy.

%
%

\subsection{Results on Fe partitioning}

%
%
\begin{figure}
\includegraphics[width=1.00\textwidth]{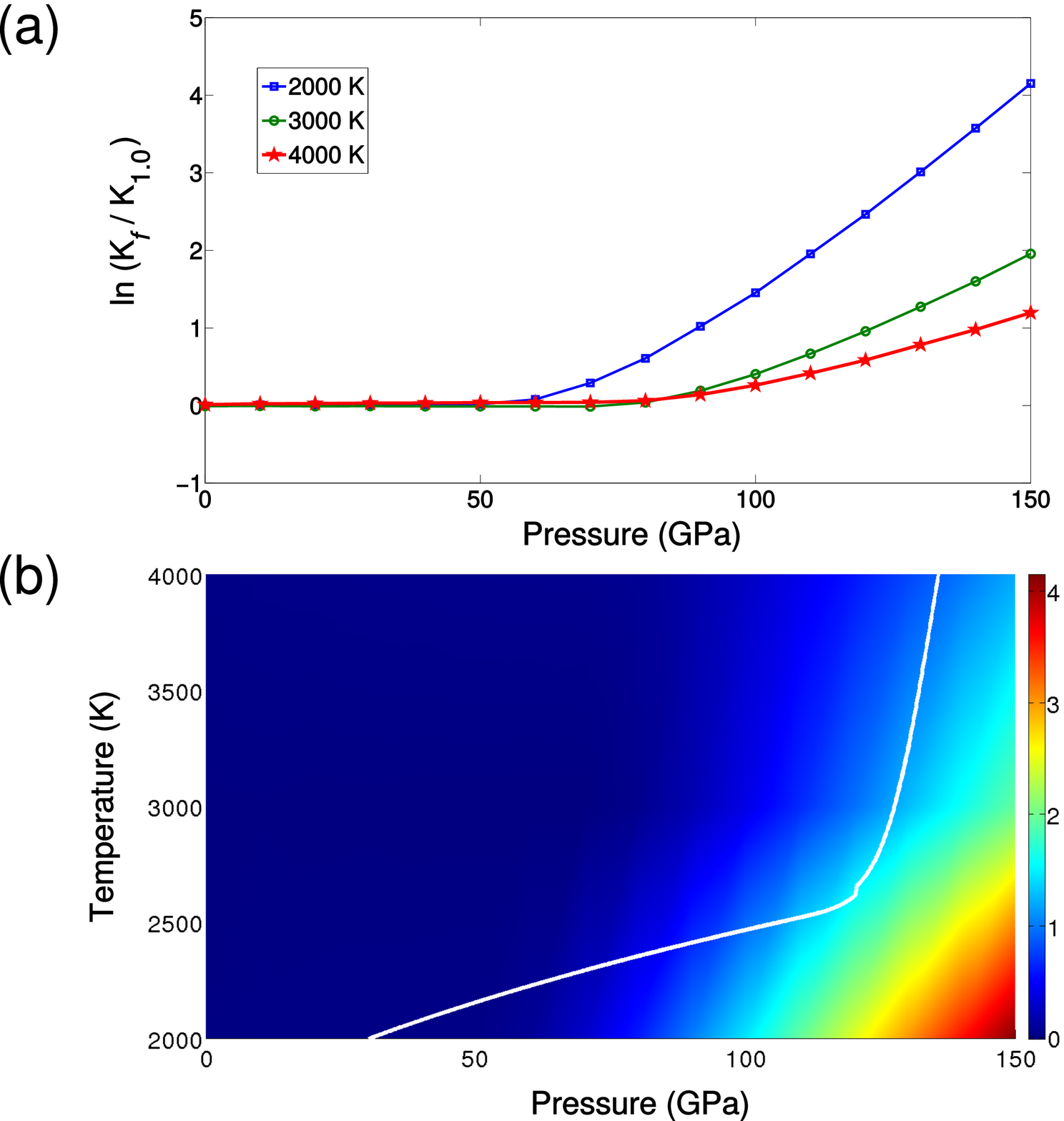}
\caption{a) Ratio of partition coefficient of Fe between
  ferropericlase and perovskite assuming the equilibrium high-spin
  fraction ($K_{f}$) and all Fe in the high-spin state ($K_{1.0}$). b)
  Bilinear interpolation of the results of a), along with a
  geotherm~\cite{Stixrude2009} (white solid
  line). \label{fig:thermodynamic_panel}}
\end{figure}

Results for the ratio of the two partition coefficients described in
the main article are shown in Figs~\ref{fig:thermodynamic_panel}a and
b.

\subsection{Electronic density of states and metallization}

\begin{figure}
\includegraphics[width=1.00\textwidth]{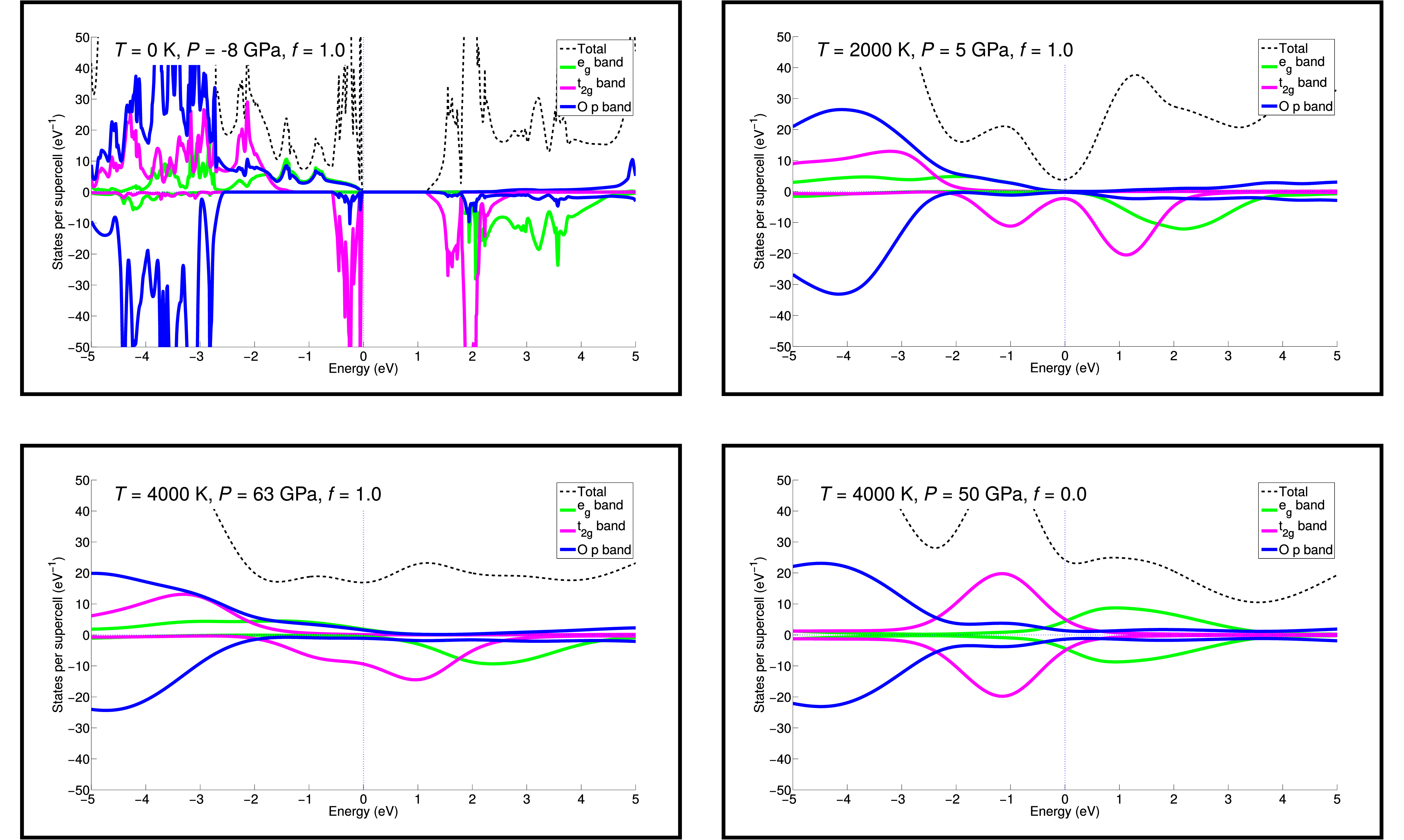}
\caption{Total and projected DOS of ferropericlase illustrating the
  effect of pressure and temperature as well as high-spin fraction on
  the DOS near the Fermi level. The plots are shifted so that the
  Fermi level is at 0~eV.\label{fig:dos_panel}}
\end{figure}

In Fig.~\ref{fig:dos_panel}, we show the electronic density of states
(DOS) of ferropericlase. Increasing temperature and pressure causes
the DOS near the Fermi level to increase significantly, and hence
electrical conductivity to increase, eventually up to semi-metallic
values. Collapsing the Fe moments to the low-spin state, as during the
spin crossover, further increases the DOS near the Fermi level.


%
%
